\documentclass[12pt]{article}

\newcommand{\preprint}[1]{\begin{table}[t]  
             \begin{flushright}               
             {#1}                             
             \end{flushright}                 
             \end{table}}                     
\renewcommand{\title}[1]{\vbox{\center\LARGE{#1}}\vspace{5mm}}
\renewcommand{\author}[1]{\vbox{\center#1}\vspace{5mm}}
\newcommand{\address}[1]{\vbox{\center\em#1}}

\usepackage{amsmath,amssymb,amsfonts,
graphicx, cite}
\usepackage{dcolumn}

\usepackage[usenames,dvipsnames]{xcolor}

\usepackage[pdftex, bookmarks={false}, pdfauthor={John McGreevy}, pdftitle={all-fermion-gauge-theories}]{hyperref}
\hypersetup{colorlinks=true, linkcolor=BrickRed, citecolor=Violet, filecolor=OliveGreen, urlcolor=RoyalBlue, filebordercolor={.8 .8 1}, urlbordercolor={.8 .8 0}}
\usepackage{soul}
\setstcolor{Red}

\usepackage[normalem]{ulem}
\usepackage{enumerate}
\usepackage{epsfig}
\usepackage{yfonts}

\usepackage{bm}

\usepackage{wrapfig}

\definecolor{darkgreen}{rgb}{0,0.4,0}
\definecolor{darkred}{rgb}{0.4,0,0}
\definecolor{darkblue}{rgb}{0,0,0.4}
\definecolor{lightblue}{rgb}{.6,.6,0.9}

\def\redt{\textcolor{darkred}}

\definecolor{uglybrown}{rgb}{0.8,  0.7,  0.5}

\definecolor{davecolor}{rgb}{0.95,  0.5,  0.2}

\definecolor{celebrationcolor}{rgb}{0.75,  0.0,  0.9}

\def\eg{{\it e.g. }}
\def\ie{{\it i.e.}}

\def\({\left(}
\def\){\right)}
\def\[{\left[}
\def\]{\right]}
\def\<{\langle}
\def\>{\rangle}

\def\dd{\text{d}}

\def\CA{{\cal A}}

\def\CD{{\cal D}}
\def\CE{{\cal E}}
\def\CF{{\cal F}}

\def\CI{{\cal I}}

\def\CT{{\cal T}}

\newcommand\half{{\ensuremath{\frac{1}{2}}}}

\newcommand\ket[1]{\ensuremath{\lvert{#1}\rangle}}
\newcommand\bra[1]{\ensuremath{\langle{#1}\rvert}}

\newcommand{\HH}{\field{H}}

\newcommand{\beq}{\begin{equation}}
\newcommand{\eeq}{\end{equation}}

\newcommand{\be}{\begin{equation}}
\newcommand{\ee}{\end{equation}}
\newcommand{\bea}{\begin{eqnarray}}
\newcommand{\eea}{\end{eqnarray}}
\newcommand{\bwt}{\begin{widetext}}
\newcommand{\ewt}{\end{widetext}}

\newcommand{\bi}{\begin{itemize}}
\newcommand{\ei}{\end{itemize}}
\newcommand{\ben}{\begin{enumerate}}
\newcommand{\een}{\end{enumerate}}
\newcommand{\bca}{\begin{cases}}
\newcommand{\eca}{\end{cases}}
\newcommand{\bln}{\begin{align}}
\newcommand{\eln}{\end{align}}
\newcommand{\bst}{\begin{split}}
\newcommand{\est}{\end{split}}

\newcommand{\IC}{{\mathbb C}}

\newcommand{\IP}{{\mathbb P}}

\newcommand{\IR}{{\mathbb R}}
\newcommand{\IZ}{{\mathbb Z}}

\def\Ione{\hbox{$1\hskip -1.2pt\vrule depth 0pt height 1.53ex width 0.7pt
                  \vrule depth 0pt height 0.3pt width 0.12em$}}

\def\grad{\vec \nabla}

\def\ii{{\bf i}}

\def\aa{{\bf a}}

\def\HH{{\bf H}}

\def\gG{\textsf{G}}

\def\gSU{\textsf{SU}}
\def\gSL{\textsf{SL}}
\def\gU{\textsf{U}}
\def\gS{\textsf{S}}

\def\gs{\text{gs}}

\def\grad{\vec \nabla}

\def\ii{{\bf i}}

\def\mE{\mathsf{E}}
\def\nb{{\bf b}}

\def\aa{{\bf a}}

\def\HH{{\bf H}}

\def\lsim{\mathrel{\mathstrut\smash{\ooalign{\raise2.5pt\hbox{$<$}\cr\lower2.5pt\hbox{$\sim$}}}}}
\def\gsim{\mathrel{\mathstrut\smash{\ooalign{\raise2.5pt\hbox{$>$}\cr\lower2.5pt\hbox{$\sim$}}}}}

\def\overleftrightarrow#1{\vbox{\ialign{##\crcr
     $\leftrightarrow$\crcr\noalign{\kern-0pt\nointerlineskip}
     $\hfil\displaystyle{#1}\hfil$\crcr}}}

     \def\overleftarrow#1{\vbox{\ialign{##\crcr
     $\leftarrow$\crcr\noalign{\kern-0pt\nointerlineskip}
     $\hfil\displaystyle{#1}\hfil$\crcr}}}

\def\eps{\epsilon}

\begin{document}

\begin{titlepage}

\title{All-fermion electrodynamics
\\ and \\
fermion number anomaly inflow
}

\preprint{UCSD/PTH 14-06}

\vskip 10mm

\author{S.~.M.~Kravec${}^a$, John McGreevy${}^a$ and Brian Swingle${}^{b,c}$}


\address{
${}^a$ Department of Physics,
University of California at San Diego,
La Jolla, CA 92093, USA}

%
\address{${}^b$ Department of Physics, Harvard University, Cambridge, MA 02138, USA}

\address{${}^c$Department of Physics, Stanford University, Stanford, CA 94305, USA}

 \vskip 10mm

\begin{abstract}
\vskip 2mm

We demonstrate
that $3+1$-dimensional quantum electrodynamics
with fermionic charges, fermionic monopoles, and fermionic dyons
arises at the edge of a
$4+1$-dimensional
gapped state with short-range entanglement.
This state 
cannot be adiabatically connected to a product state, 
even in the absence of any symmetry.
This provides independent evidence
for the obstruction
found by \cite{WangPotterSenthil}
to a
$3+1$-dimensional
short-distance completion
of all-fermion electrodynamics.
The nontriviality of the bulk is
demonstrated by a novel fermion number anomaly.

\end{abstract}

\vfill

September 2014

\end{titlepage}


\vfill\eject

\tableofcontents
\vfill\eject

\section{Introduction}

Symmetry protected topological states (SPTs) with symmetry group $G$ are quantum phases of matter which cannot be adiabatically connected to the trivial phase in the presence of the symmetry $G$,
but which can be adiabatically connected to the trivial phase if the symmetry $G$ is broken.
Absent any symmetry, the lack of adiabatic connection to a product state generally implies
topological order: long-ranged entanglement and topology-dependent groundstate degeneracy 
(for a review, see \eg\cite{Wen:2012hm}). However, in rare cases, short-range entangled topological states can be non-trivial in the absence of any symmetry \cite{PhysRevB.25.2185}. For example, $2+1$-dimensional chiral states are distinguished from the trivial phase by their gapless chiral edge modes which persist in the absence of any symmetry.
In this paper, we give another example of a short-range entangled bulk topological state
not protected by any symmetry.
It is made from bosons in 4+1 dimensions
and its edge hosts a version of electrodynamics where all charged objects are fermions.

Following the aforementioned examples, it is believed that such symmetry protected topological phases are characterized by their edge states, since they appear to be trivial in the bulk\footnote
{The subject is reviewed in
\cite{Turner:2013kp, Senthil:2014ooa}.}. This definition implies that the
physics which may arise at the
edge of a $d+1$-dimensional SPT\footnote{Following Sachdev's useful convention, we'll use $D=d+1$ to denote
the number of spacetime dimensions.}
(and any low-energy effective field theory description
thereof)
must have features which may not arise intrinsically,
in the absence of the extra-dimensional bulk.
That is, there must not be a local lattice model (or other regulator)
in strictly $d-1$ spatial dimensions which regulates the edge theory
and preserves all of its symmetries.  For example, there is no way to regulate a chiral edge mode in one dimension.

This realization \cite{PhysRevB.87.235122, Kravec:2013pua} implies that the
study of SPT states
may be used to identify obstructions
to symmetric regulators of quantum field theory (QFT).
In simple examples, such an obstruction can be identified
with an 't Hooft anomaly coefficient \cite{'tHooft:1979bh},
a well-known obstruction to gauging a global symmetry of a field theory.
When realized at the edge, the bulk theory cancels the anomaly
by anomaly inflow \cite{Callan:1984sa}.
However, there are examples,
particularly for discrete symmetries,
 where there is no previously-known anomaly\footnote
{Formal attempts to interpret SPT obstructions in these
terms include \cite{Wang:2014tia, Kapustin:2014gma, Kapustin:2014zva, Kapustin:2014tfa, Kapustin:2014lwa}.}.
Examples of such obstructions which go beyond familiar
global anomalies include
many interesting states in $2+1$ dimensions,
such as the algebraic vortex liquid \cite{PhysRevB.87.235122},
time-reversal-invariant $\IZ_2$ gauge theory where all quasiparticles are fermions
(the ``all-fermion toric code") \cite{PhysRevB.87.235122, Burnell:2013bka},
other topologically ordered states
in $2+1$ dimensions \cite{PhysRevB.88.115137, Metlitski:2013uqa, Bonderson:2013pla,
2013PhRvX...3d1016F, 2014arXiv1403.6491C, 2013arXiv1306.3286M},
and a simple three dimensional example
\cite{Kravec:2013pua}.

This paper may be regarded as a sequel to
\cite{Kravec:2013pua},
which identified an obstruction
to a regulator for `pure' $\gU(1)$ gauge theory
which manifestly preserves electromagnetic duality\footnote
{The edge theory of this model was studied further in
\cite{Amoretti:2014kba}.}.
While this is a gaussian model, such a no go result
is interesting given attempts to construct such manifestly duality-symmetric realizations \cite{Schwarz:1993vs}.
Further, it shows the impossibility of {\it gauging}
electromagnetic duality, a conclusion
which was argued from a very different point of view in \cite{Deser:2010it, Bunster:2011aw, Saa:2011wb}.

Here we point out that a stronger obstruction may be found by adding `matter'
to the bulk model studied in \cite{Kravec:2013pua}.
The model we find at the surface
is $3+1$-dimensional electrodynamics
where all of the minimally-charged (electrically and/or magnetically)
particles are {\it fermions}.
This system has been discussed
recently in \cite{WangPotterSenthil},
which demonstrated
that it does not admit an interface with vacuum -- it is not `edgeable'.  To be precise, ref.~\cite{WangPotterSenthil} showed that all-fermion electrodynamics cannot be realized in $3+1$-dimensions if the microscopic regulator consists entirely of bosonic degrees of freedom.  If we add to the microscopic physics gauge-invariant fermion degrees of freedom, then we can bind the gauge invariant fermion to the minimally charged fermionic objects to produce minimally charged bosonic objects.  Bosonic electrodynamics of course can be regulated in strict $3+1$-dimensions, by $\gU(1)$ lattice gauge theory \cite{Wilson:1974sk},
or (more locally) by a $\gU(1)$ toric code \cite{Kitaev:1997wr, Levin:2005vf}.

This example is particularly dramatic
because the obstruction arises in the absence of any symmetry
(including translations).
The $4+1$-dimensional phase is a short-range-entangled state
which is guaranteed to have an interesting edge
no matter what horrors of disorder we subject it to.
The only other known examples of this type are
(copies of) the fermionic chiral ($p+ip$) superfluid states\footnote{in which the $\IZ_2$ symmetry is ungauged and the vortices are not dynamical objects}, Kitaev's $\mathsf{E}_8$ state of bosons \cite{Kitaev-unpublished, Lu:2012dt, BGS-unpublished}
(both in $2+1$ dimensions), and Kitaev's majorana chain in $1+1$ dimensions \cite{Kitaev:2000} (provided we assume fermion number is unbreakable).

We note that the classification of \cite{Kapustin:2014tfa} includes a nontrivial state in $4+1$ dimensions
without symmetry.  Ref.~\cite{Thorngren:2014pza} attributes fermionic
excitations to its surface states.  We anticipate that
the independent construction in this paper can be interpreted
as a physics-based realization of the machinery in that work.

{\bf Why is the bulk nontrivial?}
The fact that the edge of the $4+1$-dimensional system
realizes all-fermion electrodynamics, combined with an argument that all-fermion electrodynamics cannot be regulated in $3+1$ dimensions, implies that the bulk is a non-trivial $4+1$-dimensional state of matter. Ref.~\cite{WangPotterSenthil} (Appendix D) has given
one such argument
for the absence of a $3+1$d regulator of all-fermion electrodynamics.
Hence the bulk is a nontrivial state of matter, any representative groundstate of which is
not adiabatically deformable to a product state.
We provide two independent demonstrations of bulk non-triviality, one
from the point of view of the edge (in \S\ref{sec:edge}),
and one that uses directly the bulk (in \S\ref{sec:fermion-anomaly}). Since no symmetry is required to define the bulk state, it is a topological phase of matter which is protected from all weak Hamiltonian perturbations. However, it is still short-range entangled 
\cite{Kitaev-2013, 2014arXiv1407.8203S}: two copies of the bulk state can be deformed into a product state, so it is its own `inverse state'.

In the context of microscopic {\it bosonic} phases, the only other known example of a short-range entangled state which is distinct from the trivial phase in the absence of any symmetry is the $\mathsf{E}_8$ state in $D=2+1$ dimensions\footnote{as well as multiple copies of this state, which comprise an integer classification}.  As stated above, the distinguishing feature of the $\mathsf{E}_8$ state is its chiral edge modes at an interface with the vacuum.  A sharp and universal characterization of these chiral edge modes is the thermal Hall response: heat will be transported uni-directionally without dissipation along the boundary of the sample.
In the language of anomalies, the nontriviality of the above $D=2+1$ example
is demonstrated by the chiral central charge
$c_- \equiv c_L-c_R$ of
the edge states.
$c_-$ represents a gravitational anomaly of the edge CFT,
and this is a construction of
gravitational anomaly inflow.

In the $4+1$-dimensional example studied here, the analogous signature of the nontriviality of the state
seems to be fermion number anomaly inflow, as we show in \S\ref{sec:fermion-anomaly}.

We demonstrate that this effect also occurs in the beyond-cohomology $D=3+1$ boson SPT
protected by time reversal symmetry studied in
\cite{PhysRevX.3.011016, PhysRevB.87.174412, PhysRevB.87.235122, Burnell:2013bka, Metlitski:2013uqa}\footnote{A related phenomenon was described for 
edge states of 3+1d SPTs whose protecting group contains $\gU(1)$
in \cite{Metlitski:2013uqa}.
In that case, the anomaly occurs upon gauging the $\gU(1)$.
}.
A possible surface termination of this SPT consists of
an all-fermion toric code, a model which has no $D=2+1$ realization with time reversal symmetry.
Our claim implies that the preservation of time-reversal in
the all-fermion toric code comes at the cost
of the conservation of fermion number!

We emphasize that the main conclusion of this paper pertains
to models made from {\it bosons} in $D=4+1$ dimensions.  As we show, the addition of microscopic gauge-invariant fermions to the system removes any obstruction to realizing the edge physics in strict $D=3+1$ dimensions.
Such a gauge-invariant local fermion cannot arise at the edge of a bosonic system.
From the point of view of a lattice field theorist
attempting to regularize the given low-energy field theory,
having to add an extra species of massive fermion at the cutoff may not
seem like a huge price.
However, we regard the demonstration that
such a step is {\it required} as fascinating
and requiring a systematic understanding.

The paper is structured as follows.  First (\S\ref{sec:bdc}), we review the physics of 
two-form Chern-Simons (``$BdC$") theory in $4+1$ dimensions and show that it admits an edge which supports all-fermion electrodynamics. The group of electromagnetic duality transformations, which can be realized as an exact symmetry of the bulk $BdC$ theory, plays an important role in the analysis.  Second (\S\ref{sec:BadThing}), by considering the path integral of all-fermion electrodynamics on $\mathbb{CP}^2$, we show that all-fermion electrodynamics cannot have a bosonic regulator.  This constitutes a proof of bulk non-triviality via edge non-regularizability.  Third (\S\ref{sec:coupled-layer}), we show how to construct the bulk non-trivial state from layers of ordinary (e.g., with bosonic charges) electrodynamics by condensing dyon strings. Finally, we show how to interpret the obstruction in terms of a fermion number anomaly of the all-fermion electrodynamics (\S\ref{sec:fermion-anomaly}) and show that similar physics is realized in the non-trivial time-reversal ($\CT$) protected bosonic SPT in $3+1$ dimensions (\S\ref{sec:all-fermion-TC}).

\section{The $BdC$ model coupled to matter}
\label{sec:bdc}

\subsection{$BdC$ summary}
We begin by describing the action of the $BdC$ theory and reviewing its basic properties 
\cite{Schwarz:1978cn, Schwarz:1979ae, Witten:1988hf, Elitzur:1989nr, Horowitz:1989ng, Blau:1989bq, Horowitz:1989km, Witten:1998wy, 
Moore:2004jv,
Belov:2004ht, Hartnoll:2006zb, Mooretalk}.  Consider $2$-forms $B^I_{MN}$ ($I=1..N_B$ labels the form, $MN$ are the spacetime indices) in
$4+1 $ dimensions, with the topological action
\def\nB{N_B}
\be \label{eq:normalization}
 S[B] = {K_{IJ}\over 2 \pi } \int_{\IR \times \Sigma}  B^I \wedge \dd B^J \ee
where $\Sigma$ denotes the space of interest.

To process this action, we need a little exterior algebra: a $p$-form $\alpha_p$ and a $q$-form $\beta_q$ satisfy $\alpha_p \wedge \beta_q = (-1)^{pq} \beta_q \wedge \alpha_p$ and $\dd(\alpha_p \wedge \beta_q) = \dd\alpha_p \wedge \beta_q + (-1)^p \alpha_p \wedge \dd\beta_q$. Hence we have $B^I \wedge B^J = B^J \wedge B^I$ and 
\be\label{eq:B-to-C}
B^I \wedge \dd B^J = \dd(B^I \wedge B^J) - \dd B^I \wedge B^J~,\ee
so up to a total derivative the action is anti-symmetric in $IJ$.  Thus $K$ is an anti-symmetric $2\nB \times 2\nB$ matrix.
Shortly we show
that in order for \eqref{eq:normalization} to govern
the low-energy effective field theory of
a short-range entangled bulk state,
$K$ must also be an integer matrix with $\det(K)=1$. Also, since $ B \wedge \dd B  = \half \dd (B \wedge B) $ is a total derivative,
we must have an even number of such two-forms.
To see that there must be an even number,
note that we view the topological field theory action \eqref{eq:normalization}
as the extreme low-energy effective field theory for a gapped state of matter;
this is only self-consistent if it is stable to perturbations by generic irrelevant bulk terms
involving these low-energy degrees of freedom.
If there were an odd number of two-forms, the addition of generic irrelevant bulk terms --
in particular the bulk Maxwell term
\be\label{eq:max5} {1\over g^2} \int_{\IR \times \Sigma} d B \wedge \star d B\ee
(where $\star$ is the Hodge duality operation)
-- would produce a propagating gapless photon in $D=4+1$.

The local gauge transformations $ B^I \simeq B^I + d \lambda^I$ are
redundancies of the model.
An important further ingredient of the definition of the model \cite{Maldacena:2001ss, Belov:2004ht, Mooretalk} is
the `large gauge' identifications:
\be\label{eq:largegauge} B^I \simeq B^I + n^\alpha \omega_\alpha , ~~~
[\omega^\alpha] \in H^2(\Sigma, \IZ), ~~ n^\alpha \in \IZ^{b^2(\Sigma)} ~,\ee
where the betti number $ b^2(\Sigma) \equiv \text{dim} H^2(\Sigma, \IZ) $
is the dimension of the second integer cohomology of $\Sigma$.
This requires the entries of $K$ to be integers\footnote{In this paper we will only discuss
this model on manifolds without torsion homology.  For the machinery required to lift
this restriction, see \cite{Freed:2006yc}.}.

The equations of motion following from (\ref{eq:normalization}) are, $\forall I$,
\be \label{eq:bdceom}
K_{IJ} d B^J = 0.
\ee
When $K$ has full rank, these equations are solved by flat two-form fields,
which are identified by local gauge equivalences,
and there are therefore no local degrees of freedom.
As a result,
the gauge-inequivalent operators
(analogs of Wilson loop operators)
are labelled by cohomology classes
\def\xxi{m}
\be\label{eq:fluxop} \CF_\omega(\xxi) \equiv e^{ 2 \pi \ii \xxi_I \int_\omega B^I} \ee
with $[\omega] \in H^2(\Sigma, \IZ)$.
The identification \eqref{eq:largegauge} on $B$ implies
$ \xxi^I \in  \IZ$.

Using equal-time canonical commutators for $B^I$, the flux
operators \eqref{eq:fluxop} satisfy a Heisenberg algebra:
\be\label{eq:heis} \CF_{\omega_\alpha}(\xxi)
\CF_{\omega_\beta}(\xxi')
= \CF_{\omega_\beta}(\xxi')
\CF_{\omega_\alpha}(\xxi) e^{ 2 \pi \ii  \xxi_I^\alpha \xxi_J^{'\beta} {\(K^{-1}\)^{IJ} \CI_{\alpha\beta} }}. \ee
Here
$$\CI_{\alpha\beta} \equiv \int_{\Sigma} \omega_\alpha \wedge \omega_\beta $$
is the intersection form on $H^2(\Sigma, \IZ)$,
which is a $b^2(\Sigma)\times b^2(\Sigma)$ symmetric matrix.
Consider the minimal case (relevant later on)
where $\Sigma = \IC\IP^2$, for which $b^2(\IC\IP^2) = 1$
and $ \CI = 1$.
The smallest representation
of the algebra \eqref{eq:heis} is then $|{\rm Pf}(K)|$-dimensional,
which must be unity for short-range-entangled states.
Hence we require $\det K = {\rm Pf}^2(K) = 1$.

The $BdC$ theory is a special case of (\ref{eq:normalization}) where we take $N_B=2$ and let $B^1 = B$, $B^2 = C$, and $K = k i \sigma^y$; we must set $k=1$ for this state to be short-range entangled (when $k>1$ the system has topological ground state degeneracy depending on $b^2(\Sigma)$, namely
$k^{b^2(\Sigma)}$ groundstates).
We now review its physics on a space with boundary
\cite{Kravec:2013pua, Amoretti:2014kba}.
In the presence of a boundary,
the solutions of the equations of motion
produce physical excitations:
a one-form field $a$ localized at the boundary.
This mode is physical because gauge transformations which are nontrivial
at the boundary do not preserve \eqref{eq:normalization}.
Boundary terms
(whose coefficients are non-universal)
produce the Maxwell action for $a$.
In particular,
the boundary condition
arising from
variation of an action with the leading irrelevant operators (\ie~the bulk Maxwell terms
\eqref{eq:max5})
is:
$$\( { k\over 2 \pi } B - { 1\over 2 g^2 } \star_4 C \)|_{\partial \Sigma_4} = 0~. $$
Upon a convenient rescaling, the identification of boundary degrees of freedom is:
\be\label{eq:field-identification}
B = d a,~~~ C = \star d a ~.\ee

An important symmetry of the topological action \eqref{eq:normalization} is the group $\gSL(2 N_B,\IZ)$ of field redefinitions that preserve the identifications \eqref{eq:largegauge}. We emphasize that this symmetry is not necessary for the 4+1 bulk to be distinct from a trivial phase; indeed, this symmetry may be broken by UV physics, but it turns out to be very convenient to analyze certain topological features of the physics assuming this symmetry holds. In the case of the $BdC$ theory, the group is $\gSL(2,\IZ)$ and it is closely related to the group of duality transformations on the boundary electrodynamics.

The action of $\gSL(2,\IZ)$ on $B,C$ is in the fundamental representation
$$
\begin{pmatrix}
B \\ C
\end{pmatrix}
\to
\mathsf{M} \begin{pmatrix}
B \\ C
\end{pmatrix}  $$ with $ \mathsf{M} \in \gSL(2, \IZ)$.
The `$\mathsf{T}$ ' transformation $\mathsf{T} =  \begin{pmatrix}
1 & 1  \\ 0 & 1
\end{pmatrix}
$
is a symmetry because $ B \wedge dB$ is a total derivative;
by
\eqref{eq:field-identification},
this transformation shifts the theta angle of the surface gauge theory by $ 2\pi$.
The `$\mathsf{S}$' transformation
$\mathsf{S} =  \begin{pmatrix}
0 & 1  \\ -1 & 0
\end{pmatrix}$ is a symmetry because of \eqref{eq:B-to-C},
and acts
as electromagnetic duality on the boundary gauge field.
These two transformations generate $\gSL(2,\IZ)$.
Notice that on $B,C$, the $\IZ_2$ center of the duality group acts nontrivially
(this is charge conjugation at the edge).

\subsection{Coupling to strings (matter)}

Just as a one-form gauge field $A$ couples minimally to the worldline of a charge, $ \int_\text{worldline} A$,
a two-form gauge field $B$ couples minimally to the worldsheet of a string, $ \int_\text{worldsheet} B$.
Adding matter to Chern-Simons theory
is usually
\cite{wen04}
described in terms of a statistics vector, $l_I$,
so that the quasiparticle (here, `quasistring') current is the two-form $ l_I \star d B^I$.
If $B^I$ are normalized as in \eqref{eq:normalization},
the $l_I$ must be integers,
so that $e^{ \ii \oint_{\Sigma} l_I B^I } $ is periodic
under shifts of the periods of $ B^I$ over all
topologically nontrivial 2-cycles $\Sigma$.

Gauge invariance under $B^I \simeq B^I + d \lambda^I$ requires that strings not end in the bulk of the sample.  However, strings can end at an interface with vacuuum. Then because of the identification \eqref{eq:field-identification},
the ends of the strings are electric and magnetic charges under the boundary gauge field $a$. Indeed, given a string which terminates at a boundary, the coupling $\int_\text{worldsheet} B$ reduces to the coupling $\int_\text{worldline} a$ by Stokes' theorem.

We discuss in detail below the statistics of the surface particles arising at the ends of the bulk string matter. As a preliminary, note that the modular group $\gSL(2, \IZ)$ acts on the string matter as well. This action is necessary to preserve the coupling between string worldsheets and two-form fields.

\subsection{Edge physics}
\label{sec:edge}

We now consider an edge of the $D=4+1$ dimensional $BdC$ bulk which supports $\gU(1)$ electrodynamics in $D=3+1$ dimensions \cite{Kravec:2013pua}.  As anticipated in the introduction, the crucial question is: what are the statistics of the basic charged particles on the edge.  Because the edge electrodynamics is a stable phase of matter and because the statistics of the charged particles is topological data, these statistics must be stable to the breaking of all symmetries in the problem.  Hence to determine the statistics we may assume extra symmetry and be confident that we have the correct statistics even if we later break the symmetry (for example by allowing the electron and monopole to have different masses) to realize the generic situation.

Thus suppose that we preserve the manifest $\gSL(2,\IZ)$ duality symmetry
of the $BdC$ theory.
Duality symmetry implies that the charge
$e$ and the monopole $m$
have the same statistics,
since they are related by the symmetry.
For $\gG = \gU(1)$, the full duality group is $ \gSL(2, \IZ)$, and it acts on the charge vector by
$ \begin{pmatrix}
q_e \\ q_m
\end{pmatrix}
\to
\underbrace{\begin{pmatrix}
a & b  \\ c & d
\end{pmatrix} }_{\in \mathsf{SL}(2,\IZ)}
\begin{pmatrix}
q_e \\ q_m
\end{pmatrix}  $.
In particular, the transformation $ (\mathsf{T^t} \mathsf{S})^{-1} $
takes the charge to the $(1,1)$ dyon $\epsilon \equiv em $.
The boundstate with these quantum numbers must therefore have the same statistics as the charge and the monopole.
Since these are particles in $3+1$ dimensions, they may be either all bosons or all fermions.



\begin{wrapfigure}{R}{0.3\textwidth}
  \vspace{-20pt}
\hskip2in
\includegraphics[width=.3\textwidth]{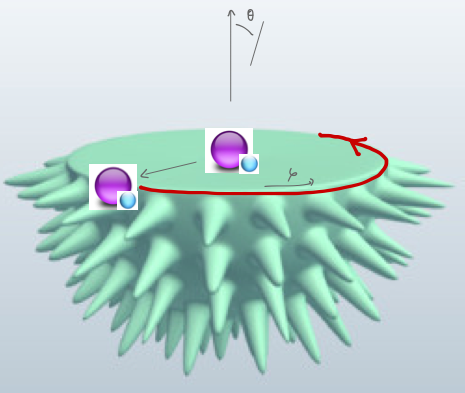}
  \caption{\small A depiction of the calculation of dyon statistics.  The spikes represent
  the flux produced by the dyon at the center.\label{fig:sod}}
  \vspace{-20pt}
\end{wrapfigure}
Naively both possibilities are allowed, but in fact, if $e$ and $m$ have the same statistics, then $\epsilon$ must be a fermion.
This phenomenon is sometimes called `spin from isospin' \cite{Jackiw:1976xx,Jackiw:1975fn}
(when the electrodynamics is UV completed by $\gSU(2)$ gauge theory with an adjoint higgs field). Note that we must assume there are no gauge-invariant fermions around, otherwise we could bind such a fermion to the dyon without changing its charges and turn it into a boson.

To see this efficiently, consider two identical dyons well-separated in space
compared to any cutoff scales.
Since they are identical particles, moving one of them adiabatically
in an arc of angle $\pi$ around the other
results in the same state (up to an innocuous center-of-mass translation).
The Berry phase acquired in doing so is
$$ \varphi = e \int_0^{\pi} d\varphi \underbrace{\CA_\varphi(\theta={\pi\over 2}, \varphi)}_{\text{Dirac
monopole field}}
\buildrel{\text{Dirac}}\over{=} \pi g e ~. $$
If $g$ and $e$ have the minimal charges, saturating the Dirac quantization condition,
then
$$ \psi(x_1, x_2) = e^{ \ii \varphi} \psi(x_2, x_1)  = - \psi(x_2, x_1)$$
and these particles are fermions.  The extra $\frac{\hbar}{2}$ unit of angular momentum comes from the electromagnetic field.  Note that any exchange phase coming from the constituent $e$ and $m$ particles cancels because we assumed they were both bosons or both fermions.

Thus we reach the remarkable conclusion that the model with a duality-symmetric spectrum of all bosons is not even self-consistent!  On the other hand, an all-fermion spectrum is self-consistent: because of the
additional $ \hbar \over 2 $ unit of angular momentum
in the electromagnetic fields,
the dyon boundstate of two fermions is still a fermion \cite{Witten:1979ey}.

To prove that the bulk is non-trivial we argue by contradiction and suppose that all-fermion electrodynamics can be realized in strict $D=3+1$ dimensions with microscopic bosons only. Then we could place a field theory realization on $\mathbb{CP}^2$ since the theory is bosonic and requires no spin structure for its definition.  However, something bad happens, which we describe next, in \S\ref{sec:BadThing}.

Hence there must be no UV completion in the same dimension with only microscopic bosons.  Since the BdC theory provides a UV completion of all-fermion electrodynamics with only bosons 
at its edge, it follows that the bulk $BdC$ phase is necessarily distinct from the trivial phase. Alternatively, the results of \cite{WangPotterSenthil} also imply that all-fermion electrodynamics cannot be realized in strict $D=3+1$ dimensions without gauge-invariant fermions, so again we conclude that the bulk $BdC$ phase is distinct from the trivial phase.

\section{The Bad Thing that Happens on $\IC\IP^2$}
\label{sec:BadThing}
To show the impossibility of a bosonic regulator of all-fermion QED, we show that there is no consistent way to define the partition function on $\IC\IP^2$.  To make the argument we suppose:\\
{\bf Postulate 1}:
A $\gU(1)$ gauge theory with gapped matter
(and hence the value of the $\gU(1)$ gauge theory path integral on a closed manifold M, modulo non-universal garbage)
is specified by the theta angle and the coupling
and by the spectrum of charges.

But what theta angle and coupling you ask?  What data about the spectrum?  More specifically, we suppose:\\
{\bf Improved Postulate 1}: The value of the gauge theory path integral on a closed manifold $M$, modulo non-universal garbage, depends only on the bare coupling $\tau = \theta + {4\pi \ii \over g^2}$
(and $\bar \tau$), and on a choice of statistics for
the excitations with minimal electric and magnetic charges,
$e$, $m$.
We include the dependence on the particle masses and various other couplings
in the category of `non-universal garbage'.

A crucial point here is that
the effective theta angle (at energies below the gap to charged excitations)
may receive contributions from integrating out the matter,
as is familiar from the study of topological insulators (\eg\cite{Maciejko:2010tx, Swingle:2010rf}).
A useful perspective then, is that all such gauge theories may be realized by coupling
``pure" $\gU(1)$ gauge theory to bosonic or fermionic matter in various $\gU(1)$ SPTs,
that is: gauging the $\gU(1)$ global symmetry.
A possibility which we must also discuss
is a case with {\it no} charged matter, studied
with related intent in \cite{Witten:1995gf, Kravec:2013pua}.

Let us consider the action of duality on the gauge theory partition function.
We are free to relabel the gauge fields using the electric-magnetic duality group
$\tau \to { a \tau + b \over c \tau + d }$, ($a,b,c,d \in \IZ, ad-bc = 1$) but we must keep track of the particle statistics as well.
We will be most interested in the $\mathsf{T}$ transformation which takes $ \theta \to \theta + 2\pi$.
Recall \cite{Witten:1979ey} that shifting the theta angle
produces a spectral flow on the charge lattice:
monopoles acquire electric charge proportional to $ \theta \over 2\pi$.

Therefore (in the absence of other data, an absence for which we argue below)
the choice of statistics of the charged matter gives an invariant meaning to the duality frame.
Denote the statistics labels on the gauge theory as follows:
BBF if $e$ is a boson, $m$ is a boson and (therefore) $em$ is a fermion, BFB if $e$ is a boson, $m$ is a fermion, $em$
is a boson, etc.
Note that by the spin-from-isospin argument, this labeling is redundant (the
statistics of $em$ is determined by those of $e$ and $m$), but it will help emphasize the important distinction
between the all-fermion case and the other cases.
If we allow neutral fermions, then we
have both bosons and fermions in each charge sector, and the labeling scheme breaks down;
we assume no neutral fermions.
If there are no charged particles, then any duality transformation in $\gSL(2,\IZ)$ is a
redundancy: a relabeling of fields.

\def\gT{\mathsf{T}}
For example, the Witten effect \cite{Witten:1979ey} implies that
for any four-manifold $M$,
$$\gT: Z_M(\tau,BBF) = Z_M(\tau+1,BFB).$$
On the other hand, consider the case where $M = \IC\IP^2$;
this example is interesting because it has a two-cycle $h$ with unit self-intersection.
This means that a line bundle with ${\bf c}_1 = h$ has
$$ {1 \over 4 \pi^2} \int_{\IC\IP^2} F\wedge F  =  1 .$$
Therefore the partition sum is
$$ Z_{\IC\IP^2}(\theta) = \int DA e^{ - S_0[A]+  i {\theta \over 8 \pi^2} \int_{\IC\IP^2} F\wedge F } = \sum_{{\bf c}_1 = n h }
\int_{C_n} [DA]_n e^{- S_0[A]+  i { \theta \over 2 } n }$$
where $C_n$ labels the sector of the gauge field configuration space with $ \int_h { F \over 2 \pi } = n$.
$Z_{\IC\IP^2}(\theta)$ is therefore periodic in $\theta$ with period $ 4 \pi$.
(This fact is discussed in detail in \cite{Witten:1995gf};
the odd intersection form on $\IC\IP^2$
also plays a role in the discussion of \cite{Kapustin:2014tfa}.)

Since we know that $Z_{\IC\IP^2}(\tau, BBF)$ is not the same as
$Z_{\IC\IP^2}(\tau+1, BBF)$
 but that it is the same as
 $Z_{\IC\IP^2}(\tau+2, BBF)$,
 it follows that integrating
 out charged matter which makes the monopole a fermion generates an extra theta term
 with coefficient $2 \pi$ (mod $4 \pi$), in agreement with previous results \cite{Witten:1979ey}.

Finally, let us turn to the case of $Z(\tau,FFF)$.  By the Improved Postulate 1 we have
$$Z_M(\tau,FFF) = Z_M(\tau+1,FFF)$$
for all 4-manifolds $M$ on which the theory is defined.
However, this equation can only be true if $M$ has an even intersection form.
If the theory had a bosonic regulator, then we could place it on manifolds with an odd intersection form and no spin structure, such as $\IC\IP^2$.  The theory cannot be placed on manifolds with odd intersection form, hence the theory does not have a bosonic regulator\footnote{Note than an additional consequence of its lack of spin structure is that $\IC\IP^2$ cannot occur as the boundary of some smooth, compact $5$-manifold; it has a non-vanishing Stiefel-Whitney number. See Theorem 4.10 of \cite{milnor1974characteristic}.
This theorem prevents a contradiction with the fact that the partition function of the all-fermion electrodynamics
on $M$ {\it can} be obtained from the $BdC$ theory on a space whose boundary is $M$.
Two disjoint copies of $\IC\IP^2$ can occur as the boundary of {\it e.g.}~$\IC\IP^2 \times [0,1]$.
In this case,
the instanton sums in the two copies of all-fermion electrodynamics are correlated by the fact that
$ B = da_1 + da_2 $ is flat in the bulk, again avoiding contradiction.}.

In order for this periodicity in $\theta$
$$Z_{\IC\IP^2}(\tau + 1, FFF) \buildrel{!}\over{=} Z_{\IC\IP^2}(\tau, FFF)$$
to be a consistency condition
(that is: its violation is a gauge anomaly)
we require that
the modular properties of the partition function are
determined entirely by the spectrum of electric and magnetic charges.
We argue for this claim in a series of comments, which can be regarded
as an attempt to make precise the lack of structure in $\gU(1)$ gauge theory:

\begin{itemize}
\item
First, we emphasize that the statistics of particles
in {\it all} charge sectors $(q_e, q_m)$ are fixed
by the elementary ones (1,0), (0,1) (the generators of the charge lattice)
and the demand that there are no neutral fermions.
For example, the spectrum of the $FFF$ theory
cannot contain a magnetic-charge-two monopole which is a fermion,
because then binding such an object to the (boson) boundstate of two charge (-1) monopoles would produce a neutral fermion.

\item
In gauge theories with more interesting gauge group or massless matter content,
other labels are required to specify the partition function.
For example, gauge theories where a $2\pi$-shift of $\theta$
produces a different gauge theory
were discussed recently in
\cite{Aharony:2013hda}.
The new labels there arise
from extra topological invariants (beyond the Pontryagin invariant)
of gauge bundles whose structure group
(the gauge group)
is semisimple but not simply connected (a pedagogical exposition
of this subject can be found in \S3 of \cite{Vafa:1994tf}).

\item
Here we are studying $\gG = \gU(1)$ where this issue does not arise.
That is: The smooth topological data of a line bundle
(the structure group is $\gU(1)$)
on a simply connected manifold
is just the first Chern class (for a discussion which makes this clear see {\it e.g.}~page 3 of \cite{Witten:1995gf}).
Therefore this possibility for modifying the periodicity of theta is not available.

\item
Another potential source of a theta-dependent phase in the partition function
is a possible $\tau$-dependence in the gravitational couplings in the effective action
for the gauge fields upon integrating out the gapped charged matter.
Such couplings are crucial in computing
the partition function
of topologically twisted gauge theories \cite{Vafa:1994tf} on various four-manifolds,
and are discussed further in \cite{Witten:1995gf}.
In that context, such terms produce anomalous factors under the $\gS$ transformation,
but not under the $\gT$ transformation.

Further, to see that this is not a meaningful loophole here, we can take the
perspective described above: we couple an SPT with $\gG = \gU(1)$ symmetry
(in curved space)
to the electromagnetic field.
The gravitational effective action for the SPT is
completely fixed before the coupling to the EM fields,
which is when $\tau$ is introduced.
Therefore, the $\tau$ dependence of the action below the gap
is completely fixed by the matter content.

So the basic question is: what other kinds of UV gerbils can there be in $\gU(1)$ gauge theory
which might affect the $\tau$-dependence of the partition function?
We can see that the answer is `none' as follows.

\item
Adding fermions restores
$2\pi$ periodicity of the theta angle.
This matches nicely with the
fact
\cite{Senthil:2012tm, Vishwanath:2012tq}
that
the $\theta$ angle for a {\it background} gauge field
is only periodic mod $4\pi$ in a system made of bosons
(since the surface at $ \theta = 2\pi$ would have
odd-integer quantum Hall response, which
is not compatible with bosonic statistics of all neutral excitations).
This argument implies
that {\it only} fermions in the charge spectrum can change the periodicity in theta by $2\pi$.
But we've already accounted for the fermionic charges.

\item
As a nice corroboration of our understanding, note that the counting of non-trivial states here is consistent with the counting of SPT states \cite{Vishwanath:2012tq}.
In particular, absent time reversal, the three states BBF, FBB, and BFB are smoothly connected.

\item Finally, we believe that the argument described here
implies that {\it there is no such thing as `pure' $\gU(1)$ gauge theory},
\ie~$\gU(1)$ gauge theory without any charged matter at all.
From the low energy point of view, the problem with the all-fermion model
is simply that the spectrum is duality invariant, and so cannot be rearranged by the Witten effect.
The same is true if there are {\it no} charges, and so we have:
$$Z_{M}(\tau + 1, ---) \buildrel{!}\over{=} Z_{M}(\tau, ---)$$
(where the dashes emphasize the absence of charged matter).
The fact that this demand is violated for $M = \IC\IP^2$ was observed in
\cite{Witten:1995gf}.  We believe that the above argument
implies that this failure should be regarded as an inconsistency.
We note that there is no known regulator of this model.
The $\gU(1)$ toric code is described at low energies
by electromagnetism coupled to gapped matter with spectrum BBF.
Ordinary lattice gauge theory is simply the limit of the toric code
where the electric excitations are made infinitely heavy;
in particular it still contains gapped magnetic monopole excitations.
(A term by which one might try to lift these excitations completely,
{\it e.g.}~$ \sum_\text{plaquettes} \Delta \cdot \( \Delta \times a\)  $,
is not single-valued under the equivalence $ a_\ell \to a_\ell + 2 \pi n_\ell, n_{\ell} \in \IZ $.)

Perhaps there exists a consistent low energy theory
where there are only magnetic charges; in that case, we have
the condition
$$ Z_M(\tau+1,-B-) \buildrel{!}\over{=} Z_M(\tau,--B) $$
which is not falsified by the lack of a spin structure of $M$.

The fact that there is an obstruction to a {\it duality-invariant} regulator
of `pure' electromagnetism
was argued in \cite{Kravec:2013pua}
(with hindsight, this result also follows from the calculation of \cite{Witten:1995gf}).
Here we are making the further claim that there is no regulator at all.
The argument above shows that there is no {\it bosonic} regulator.
Many of the other anomalies discussed in this paper may be cured
by adding neutral fermions.  In this case, it is difficult to see how
the addition of gapped, neutral fermionic excitations can help.
In particular, the fact that the fermion is neutral
means that integrating it out does not generate a theta term.
However, the presence of microscopic neutral fermions amounts to a
refusal to put the system on a manifold without spin structure, such as $\mathbb{CP}^2$!
(Since the fermions are neutral, the existence of a spin${}_c$ structure
does not help.)  So indeed there is no obstruction to a 
fermionic regulator.

\end{itemize}

We discuss below in \S\ref{sec:all-fermion-TC}
the consequences of the analogous line of argument
for the all-fermion toric code in $D=2+1$.

\section{Coupled Layer Construction}
\label{sec:coupled-layer}

In this section, we describe a 4+1d local lattice model which realizes the continuum model above,
using a coupled layer construction
(precedents for such an approach include \cite{Chalker-Coddington1988,
Hosur-Ryu-Vishwanath2010, Lu:2012dt, PhysRevB.87.235122, 2014arXiv1403.0953N}).  Like the edge-based proof of bulk non-triviality, the motivation for the layer construction comes from edge physics.  If SPTs are only non-trivial because of their edge states, then we should be able to construct interesting SPTs by sewing together pairs of edge states as follows.  

First, observe that every short-range entangled state with a non-trivial edge has an inverse short-range entangled state (obtained be reversing the orientation) with a non-trivial edge and with the property that the composite short-range entangled state has a trivial edge.  In other words, for every non-trivial (anomalous) edge $\CE$ there is another non-trivial edge $\CE^{-1}$ such that $\CE \times \CE^{-1} \sim 1$ is trivial.  We then imagine a stack of such edges, $(\CE_1 \CE_1^{-1}) ... (\CE_n \CE_n^{-1})$, which can clearly be reduced to a trivial state by pairing $\CE_i$ with $\CE^{-1}_i$.  However, we may also pair $\CE^{-1}_i$ with $\CE_{i+1}$ in such a way that the edges $\CE_1$ and $\CE_n^{-1}$ are left un-paired.  Assuming interactions are local in the layer index $n$, these remaining actual edge states cannot be paired with each other and we have produced a non-trivial bulk state.  More generally, we may take any lower dimensional ``layers" and try to couple them in a similar non-integrable fashion to produce a bulk short-range entangled state with non-trivial edge states.

\begin{wrapfigure}{R}{0.3\textwidth}
  \vspace{-25pt}
\hskip2in
\includegraphics[width=.3\textwidth]{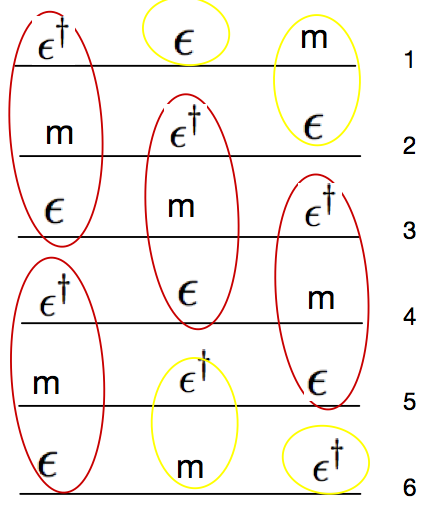}
  \caption{\small A representation of the coupled layer construction, following
  \cite{PhysRevB.87.235122}.  
The layers are coupled by condensing the objects circled in red.\label{fig:coupled-layer1}}
  \vspace{-20pt}
\end{wrapfigure}

We make a coupled-layer construction
of the all-fermion electrodynamics
 following (very directly)
the one made in \cite{PhysRevB.87.235122} for the all-fermion toric code.
It produces a trivial bosonic bulk, and the correct edge physics.
As an essential part of the construction, we are able to argue that this bosonic bulk
is well-described by the $BdC$ theory.

The method by which we
construct the bulk
can be called `dyon string condensation'.
It has a lot in common with the
dyon condensation
mechanism of statistics transmutation in 3+1 dimensions employed
in \cite{PhysRevB.88.035131}.
The construction can also be regarded as an oblique version of `deconstruction' of the extra dimension
\cite{ArkaniHamed:2001ca};
this will be a useful perspective for understanding the origin of the $ B \wedge d C$ term.

First we give a brief summary of the construction:

$\bullet$ Each layer, labelled $i=1..n$, is ordinary electrodyamics with bosonic charges:
the electron and monopole $e_i, m_i$ are gapped bosons.
This model is certainly regularizable in 3+1d by itself
on an ordinary Hilbert space of bosons on links and sites.
Denote the (fermionic) dyon in each layer as $ \eps_i$.

$\bullet$ $b_i \equiv \eps_i^\dagger m_{i+1} \eps_{i+2}$ are mutually-local bosons.

$\bullet$
Condensing $b_i$ (obliquely) {\it confines} $ a_{i+1}, i+1 = 2...N-1 $.

$\bullet$ At the top layer: $ m_1 \eps_2, \eps_1^\dagger m_1 \eps_2, \eps_1^\dagger  $ survive,
are fermions, and
 are the electron, monopole and dyon of a surviving (Coulomb-phase) $\gU(1)$ gauge field.
 A similar statement pertains to the bottom layer.

In the bulk, in the continuum, we will
arrive at the claim that this is the $ BdC $ theory with gapped string matter.

\subsection{Warmup: deconstruction of lattice electrodynamics}

\label{subsec:warmup}
\begin{wrapfigure}{R}{0.305\textwidth}
  \vspace{-30pt}
\hskip2in
\includegraphics[width=.25\textwidth]{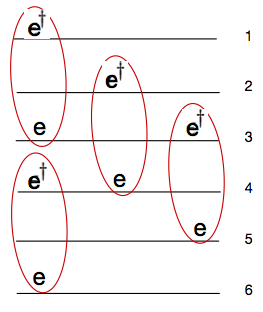}\\
\includegraphics[width=.3\textwidth]{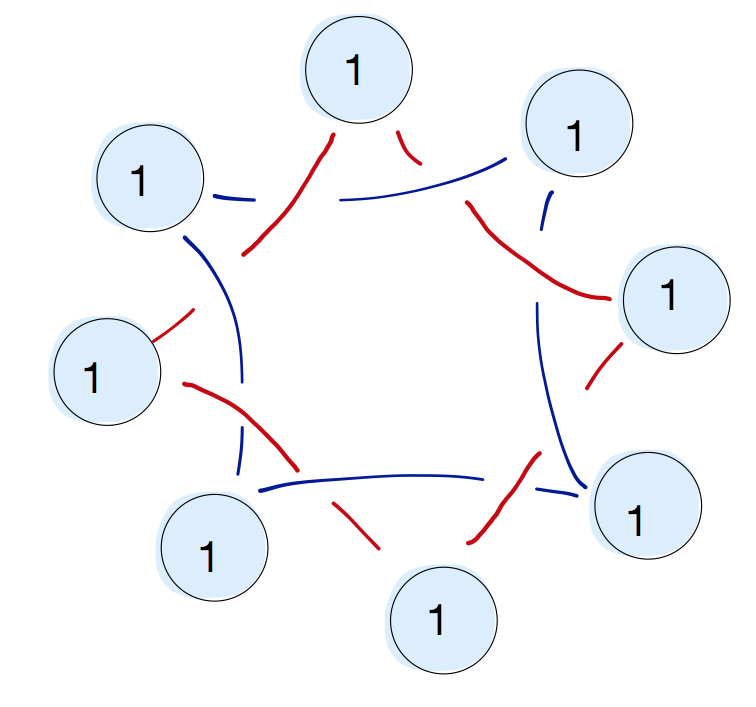}
  \vspace{-20pt}
  \caption{\small  
  Two representations of the (warmup) coupled-layer construction for
  $D=4+1$ Maxwell theory with gauge group $\gU(1)_o \times \gU(1)_e$.
  The top figure is the direct analog of the previous figure;
  the bottom is a `quiver' or `moose' diagram familiar from the high energy physics literature.\label{fig:quiver}}
  \vspace{8pt}
\end{wrapfigure}
First consider the following toy example,
which really is `deconstruction' of $4+1$d $\gU(1) \times \gU(1)$ gauge theory
on an interval, in the sense of \cite{ArkaniHamed:2001ca}.
(A quiver diagram for this construction, more familiar in the high-energy theory literature, appears in Fig.~\ref{fig:quiver}.)
Collocate an even number $N$ of layers of (cubic, say) $3d$ lattices
each of which hosts $\gU(1)$ lattice gauge theory
coupled to charge-1 lattice bosons $e_i$, with arbitrary
hopping terms in the three spatial dimensions.
For definiteness,
we could consider each layer in the zero-correlation length limit
where it is described by a solvable Kitaev-like model
with a rotor on each (oriented) link, $ [ \mE_l, \aa_{l'} ] = \ii \delta_{l, l'} $, $ \mE \in \IZ, \aa \simeq \aa+ 2 \pi $, with
$$\HH_\text{layer} = \sum_{+} \( \Delta \cdot \mE  \)^2  - \sum_{\Box} \cos  \( \Delta \times \aa \) . $$
$\Delta$ is a lattice gradient operator.
The charged bosonic matter arises at sites where $0 \neq  \Delta \cdot \mE   \in \IZ $.

Couple together the layers by the (completely local and gauge invariant) terms
\be\label{eq:condenser}
\Delta H = V \sum_x \sum_i (|\tilde \nb_i(x)|^2 - v^2)^2~.
 \ee
 Here $x$ labels a site of the 3d lattice.
 Fig.~\ref{fig:quiver} shows the case of $N=6$ layers, 
 with $\tilde \nb_i, i=1..4$ circled.
Minimizing the potential 
\eqref{eq:condenser} causes $\tilde \nb_i$ to condense,
\be  \tilde \nb_i \equiv e_i^\dagger e_{i+2}
= v e^{ \ii a_{i, i+2}}~,\ee
 higgsing $\prod_i \gU(1)_i \to \gU(1)_\text{even} \times \gU(1)_\text{odd} $.
The phases $a_{i,i+2}$ provide the link variables
in the extra dimension.
Layers with odd $i$ and even $i$ are decoupled.
The result is
$4+1d$ Maxwell theory with $ \gG = \gU(1)_\text{even} \times \gU(1)_\text{odd} $,
with massless bulk photons.
So this is not the bulk state we are looking for, but
it will be instructive.

$\gU(1)$ lattice theory in 4+1 dimensions should have
a kinetic term for the link variables along the extra dimension.
This $\mE_{x, x+\hat 4}^2$ term arises as follows.  The conjugate variable $\mE$ to $\aa$ arises from the amplitude
fluctuations of $\tilde b$:
$$ \tilde \nb_l = e^{ \ii \aa_l}( v + \mE_l),~~~~~
\tilde \nb^\dagger_l = (v+\mE_l )  e^{ - \ii \aa_l}~.$$
$$
 ~~~[\tilde \nb_i^\dagger(x), \tilde \nb_j(y) ] = - \ii \delta_{xy}\delta_{ij}
\implies [ \aa_l, \mE_{l'}] = - \ii \delta_{l l'} . $$
Expanding the condenser term \eqref{eq:condenser} about the minimum,
$\tilde \nb^\dagger \tilde \nb - v^2 = 2v  \mE + ..$, we find
$$ \Delta H = V 4 v^2\sum_{l} \mE_l^2 + .. $$

The Hamiltonian should also contain terms which suppress flux through plaquettes parallel to the extra dimension:  $ \sum_\text{plaquettes $\parallel x^4$} \cos \Delta \times \aa $.
These terms arise from microscopic gauge invariant terms including the hopping term for $\tilde \nb$:
\be
\Delta_2 H = - V_2 \sum_{x,i} \sum_{\hat\mu \neq \hat{4}} \(
\begin{matrix}
\tilde \nb_i(x+\hat\mu) 
e^{\ii \int_x^{x+\hat\mu}\aa_i - \ii \int_x^{x+\hat\mu} \aa_{i+2}}  \tilde \nb_i^\dagger(x) 
\cr + h.c.
\end{matrix}\)
\nonumber
\ee
where $\aa_i$ is the pre-existing gauge field within layer $i$. Upon condensing the $\tilde \nb_i$, the new interlayer gauge field $\aa_{i,i+2}$ combines with the existing within-layer gauge fields to form a closed Wilson loop in the $\mu 4$ plane for each term in the $\mu$ sum.

It will be useful to remind ourselves about magnetic monopoles in $\gU(1)$ lattice gauge theory
({\it e.g.}~\cite{Banks:1977cc}).
A region $R$ of the lattice
whose boundary $\partial R$ has  $ \oint_{\partial R} B  = 2 \pi g$
contains $g$ magnetic monopoles, $g\in \IZ$.
This means that the number of monopoles is not conserved on the lattice;
for example, consider a region which is a single 3-cell $V$ of the lattice;
we may change $ \oint_{\partial V} B $ from $0$ to $ 2\pi$
without changing anything,
since the gauge field is periodic $ \aa \simeq \aa + 2 \pi $
and $ B = \grad \times \aa$.

To make contact with the $BdC$ theory, it will be illuminating to dualize the odd/even
gauge fields $a^{o/e}$ to 2-form potentials: $ f^{o/e} = d a^{o/e} = \star d C^{o/e} $.
The action is
$$ S = \sum_{\alpha = o,e}  \int_{5d} \(
{1\over g_\alpha^2 }  d C^\alpha \wedge \star d C^\alpha + C^\alpha \wedge \star  j_m^\alpha \). $$

By the Meissner effect,
magnetic flux tubes of the broken relative $\gU(1)$s collimate the monopoles into {\it monopole strings}.
They must do so, since, by construction, objects magnetically charged under $a^{e/o}$
are minimally coupled to the dual field $C^{e/o}$ and must be strings.
States where the total magnetic charge in different layers is not equal
do not have finite energy.
We sequester a few more details about this to appendix \S\ref{app:vortex-sheet}.

\subsection{Dyon string condensation in more detail}

\label{subsec:real-layers}

The actual construction of the nontrivial gapped bulk is as follows.
Again each layer is ordinary electrodynamics with bosonic charges.
We will call $ \epsilon_i \equiv e_i m_i$ the dyon in each layer, which is a fermion.
The object
$ b_i \equiv \epsilon_i^\dagger m_{i+1} \epsilon_{i+2} $
is a boson (two fermions plus one boson, and no net electric charge to produce extra statistics, equals a boson).

The objects $b_i ~(i=1..N-2)$, 
for all $i$, are mutually local (\ie~their charge vectors satisfy $ q_i e_j - q_j e_i = 0, \forall i,j = 1..N-2$)
under the total $U(1)$ (in particular, they all have $q^{Total}_e = 0, q^{Total}_m =1$).
This means that it is possible to couple the layers so that these objects condense \cite{'tHooft:1977hy, 'tHooft:1979uj, 'tHooft:1981ht}.

Explicitly, we can cause them to condense by adding the
completely local gauge invariant hamiltonian
 $ \Delta H = V \sum_x \sum_i (|b_i(x)|^2 - v^2)^2$.
 The phase of the condensate $ b_i(x) = v e^{ \ii a_{i, i+2} } $
 is again a link variable along the extra dimension;
 unlike the simple construction of \S\ref{subsec:warmup}, the duality frame 
 in which this object is the vector potential rotates as we increase $i$.

 Condensing $b_i$ (obliquely) {\it confines} the gauge fields in the layers $ a_{i+1}, i+1 = 2...N-1 $.
 Objects which are not mutually local with $b_i$ are confined.
What's left?
We are condensing $N-2$ objects in a theory with gauge group $\gU(1)^N$,
so two gauge fields remain massless.
The charged objects which are mutually local with the condensate
and therefore not confined \cite{'tHooft:1977hy, 'tHooft:1979uj, 'tHooft:1981ht}
are (just as in the 2d $\IZ_2$ case \cite{PhysRevB.87.235122}):\\
$\bullet$ At the top layer : $ \epsilon_1, m_1 \epsilon_2^\dagger $ and their boundstate $ \epsilon_1 m_1 \epsilon_2^\dagger$
(and powers and products of these)
and \\
$ \bullet$ At the bottom layer:
$ \epsilon_N, m_{N-1} \epsilon_N^\dagger , \epsilon_N m_{N-1} \epsilon_N^\dagger $ etc.

At the top layer, the objects  $\epsilon_1, m_1 \epsilon_2^\dagger $ are both fermions,
and have charge $ (q_e, q_m) = (1,1) $ and $ (-1, 0)$ respectively.
The boundstate has charge $ (0, 1)$ and is therefore also a fermion, by the standard argument reviewed above,
because there is still a Maxwell field at the top layer.


\begin{wrapfigure}[7]{r}{0.3\textwidth}
  \vspace{-20pt}
\hskip2in
\includegraphics[width=.3\textwidth]{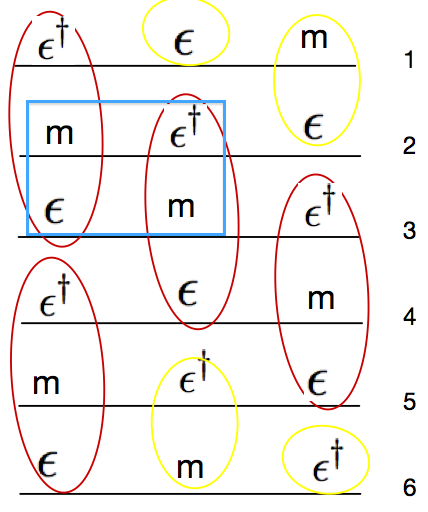}
  \vspace{-20pt}
\end{wrapfigure}
To see the full effect of condensing $b_i$, consider the blue box in the figure at right.
Although $ \eps^\dagger_i m_{i+1}$
is mutually local with $ m_i \eps_{i+1}$,
the constituents are not.
This has the consequence that
condensing $b_i$
{\it binds} the monopole strings of $a^{e/o}$
to electric flux lines of $a^{o/e}$!
This is precisely the effect of the additional term
$$\Delta S =  \int {1 \over 2 \pi} C^{e} \wedge  d C^{o}
\equiv \int {1\over 2 \pi } B \wedge d C ~~~~~~~~~~~~
$$
in the low-energy description.

\subsection{Alternative description of layer construction}

Here we make contact between the coupled layer construction
of the previous subsection and the general description
(in the section introduction)
in terms of coupled layers of $\CE$ and $\CE^{-1}$
which guarantees the correct edge states.

Again let $\CE$ denote a single copy of all-fermion electrodynamics.
First we note that the all-fermion electrodynamics is its own
inverse: $\CE = \CE^{-1}$
in the sense that
two copies of all-fermion electrodynamics can be regularized in $3+1$ dimensions.
More specifically,
$\CE \times \CE$ is deformable (by adding local, gauge-invariant interactions)
to ordinary bosonic U(1) gauge theory.
To see this\footnote{An essentially identical argument
shows that the all-fermion toric code is its own inverse.},
let $e $ and $\tilde e$ denote the electrons in $\CE$ and $ \CE^{-1}$.
Define $ b = e\tilde e^\dagger $, which is a boson.
If we condense this boson, we higgs $U(1)_{\CE} \times U(1)_{\CE^{-1}} $
to the diagonal $\gU(1)$ subgroup.
The object $e$ is a fermion charged under this gauge group;
it is related to $ \tilde e$ by taking charges from the condensate.
We should think of this object as the dyon of ordinary BBF electrodynamics,
because all of the other excitations which are mutually local with the condensate are bosons:
\begin{itemize}
\item $m, \tilde m, \epsilon= em$ and $\tilde \epsilon = \tilde e\tilde m$
are non-local with respect to the condensate, so they are confined.
\item $M \equiv m \tilde m^\dagger $ is a boson which
differs from $e$ by one unit of electric charge,
and so we should think of it as the monopole.
It is related by taking stuff from the condensate to
 $ \epsilon \tilde \epsilon $.
\item Adding $e$ to $M$ we get another boson
(since we are combining two mutually non-local fermions)
$\epsilon \tilde m^\dagger $;
apparently we should regard this as the elementary electrically charged boson.
\end{itemize}
We conclude that $\CE  \times \CE^{-1}$ is separated by simple Higgs transition from the phase $U(1)_{FBB}$, with a propagating photon
(if it's in the deconfined phase), and therefore has a $D=3+1$ regulator.

It is important to note that the remaining electrodynamics still has charged matter
which may be condensed to higgs or confine the photon;
the choice of whom to condense means that various bulk models are possible.

So, while a single copy of all-fermion electrodynamics cannot be regulated in $3+1$ dimensions, a pair of copies can be so regulated since $U(1)_{FBB}$ can be so regulated and $\CE  \times \CE^{-1} \sim U(1)_{FBB}$. The layer construction in \S\ref{subsec:real-layers}, when applied a slab of finite thickness, provides just such a regulator. As long as the thickness of the slab is not taken to infinity, the two copies of all-fermion electrodynamics can be regarded as living in $3+1$ dimensions.

Further insight into the layer construction is obtained by viewing the construction in terms of a stack of such slabs, where each slab, denoted $(\CE \CE^{-1})$, hosts two copies of all-fermion electrodynamics, one on the bottom surface and one on the top surface. The stack of slabs is denoted $(\CE_1 \CE_1^{-1}) ... (\CE_n \CE_n^{-1})$ where $i=1,...,n$ indicates the extra spatial dimension. Pairing up the all-fermion states within each slab produces the trivial bulk state in $4+1$ dimensions. Pairing $\CE^{-1}_i$ with $\CE_{i+1}$ across neighboring slabs realizes the bulk non-trivial state. This way of thinking about the layer construction realizes the motivating idea given in the section introduction.

To be a little more explicit, condensing only $b_i =  (e_{i,\text{top}} \tilde e_{i,\text{bottom}}^\dagger)$ produces layers of ordinary $FBB$ electrodynamics, by the preceding argument.
This returns us to the starting point of the layer construction of the previous section. The slabs of $FBB$ electrodynamics can then be confined to produce a trivial bulk state. 


To produce the non-trivial bulk state, the gluing may be performed by repairing the missing condensates at the top and bottom of Fig.~\ref{fig:coupled-layer1}.
In particular, think of each pair $\CE_i \CE_i^{-1}$
as a copy of Fig.~\ref{fig:coupled-layer1}.
At the top we have fermionic charges $ \epsilon_1$ and $ m_1 \epsilon_2 $;
at the bottom we have fermionic charges $ \epsilon_{N-1}^\dagger m_{N}$ and $  \epsilon_N^\dagger$.
If we glue the bottom to the top by condensing
$$ b_{N-1} \equiv  (\epsilon_{N-1}^\dagger m_{N})   \epsilon_1 $$
and
$$b_N \equiv \epsilon_N^\dagger (m_1 \epsilon_2 ) $$
then we get the BdC theory rolled up on a circle,
\ie~the coupled layer construction has translation invariance $i \to i+1$.
And in particular, there is no photon in the bulk.

\subsection{Extension to $D=3+1$ and derivation of $BF$ theory}

The logic by which we inferred the presence of the $BdC$ coupling
from the coupled layers construction
can be applied to the original construction \cite{PhysRevB.87.235122}
 of the $D=3+1$ boson SPT state with time-reversal symmetry.
The string of magnetic excitations is a vortex line;
the mutual nonlocality of the constituents of the condensed boson
glues this vortex line to the electric flux lines of the other gauge field.
The result is that the bulk model contains a term of the form ${1\over 2\pi} B\wedge F$.
That the bulk theory admits such an effective description
is well-known \cite{PhysRevX.3.011016}.
An implication of this derivation
which has not been appreciated to our knowledge
is that the all-fermion toric code --
when realized on the surface of a bosonic SPT --
suffers a fermion-number anomaly,
as we discuss in the next section.

\section{Fermion number anomaly inflow}
\label{sec:fermion-anomaly}

We will now interpret the obstruction
studied here in terms of global anomaly inflow.
The only symmetry involved in this system
is fermion parity.
We emphasize that in the bulk there {\it are} no fermions;
however, the Jackiw-Rebbi effect
demonstrates clearly that gauge fields are capable
of carrying this quantum number.

In the following we show that the fermion number conservation on the surface
of the 4+1d short-range-entangled state constructed in the previous section
is violated by high-energy processes.

There is a precedent for such violation of fermion number by quantum gauge theory.
The Witten $\gSU(2)$ anomaly \cite{Witten:1982fp} can be regarded as an anomaly
for fermion number: in a Witten-anomalous gauge theory,
instanton events create an odd number of fermions
and hence violate fermion parity conservation;
this is not something we know how to describe with a local field theory.

In the prehistory of SPT physics,
a subset of the authors  \cite{McGreevy:2011if}
studied a system where the
Witten anomaly played a crucial role in preserving the integrity
of the classification of statistics of $3+1$d particles.
In particular,
the Witten anomaly was argued to
forbid
a gauge theory whose monopoles
carry a single majorana zero mode (which monopoles, if they could be deconfined,
would enjoy non-Abelian statistics).
That paper also described a 4+1d dimensional model
whose edge realized such a gauge theory,
and therefore could be regarded as exhibiting
`Witten anomaly inflow'.

{\bf Fermion number anomaly.}
The all-fermion electrodynamics, as it arises
on a surface of the coupled-layer construction,
exhibits crucial differences from an intrinsically 3+1-dimensional system
with a bosonic regulator.
First of all, note that
the slab geometry constructed in \S\ref{sec:coupled-layer}
harbors gauge-invariant
states with a single fermionic particle at the top layer\footnote
{Here we are assuming that the 3d geometry is noncompact, so that the flux has somewhere to go.
If the 3d spatial sections are compact, we cannot have a single string stretching from one end
of the slab to the other because of the bulk Gauss law:
\be\label{eq:gausslaw} 0 =  { \delta S \over \delta B} = \star j + d C + d \star d B {1\over g^2 } \ee
which is a 3-form.  If we integrate this over a 3d region $\Upsilon$ at fixed time and codimension 1 in space,
we get
$$ 0 = (\text{ number of strings penetrating the region, counted with orientation})  +
\int_{\partial \Upsilon} ( C + \star d B /g^2 ) .$$
The last term is the usual Gauss' law term for a 2-form potential, but the important
thing is that the dependence on the fields on the RHS of \eqref{eq:gausslaw} is a total derivative.
So if there is no boundary of $\Upsilon$
-- such as if the whole space is $T^3 \times (0,1)$ and
we choose $\Upsilon $ to be the $T^3$ at some fixed position along the interval --
then the net number of strings must be zero.
}.
Since all femionic excitations carry some gauge charge (either electric or magnetic)
-- as they must in a system with a bosonic regulator --
there is no state in a putative 3+1 dimensional realization of this form.

Further, the coupled layer construction of \S\ref{sec:coupled-layer}
 directly shows that fermion number can be transported across the extra dimension, as follows.
Consider a state with an excitation of $ \epsilon_1$, the dyon at the top layer.
This excitation can for free absorb bosons
from the condensate,
which include objects of the form $b_1= \epsilon_1^\dagger m_2 \epsilon_3 $.
Combining these two objects we get something with the quantum numbers of
$ m_2 \epsilon_3$.
This looks a bit like a bulk fermion excitation,
but this object is confined (since it is not mutually local with $b_2$, which is condensed).
Also condensed is $b_3=\epsilon_3^\dagger m_4 \epsilon_5$; adding one of these in, we get
$ m_2 m_4 \epsilon_5$.
The bottom layer (for argument, we take $N=6$ layers, as in the figure above) supports a deconfined fermion excitation $ \epsilon_5^\dagger m_6 = f_\text{bottom}$.
The condensate plus top-layer excitation $f_\text{top} = \epsilon_1$ is related to this by
$$f_\text{top} b_1 b_3   = m_2 m_4 m_6 f_\text{bottom}^\dagger.$$
With arbitrary (even) $N$, we have:
$$
  f_\text{top} b_1 b_3 ... b_{N/2} = m_2 m_4 ... m_N f_\text{bottom}^\dagger.$$
This equation is understood to be true modulo the creation of neutral excitations (which
are all bosonic, by assumption).

This strongly suggests that a monopole string ($m_2 m_4 m_6...$) (bosonic, but confined)
allows fermions to tunnel from the top layer to the bottom layer.
A quantitative statement
to this effect is that
there is a nonzero amplitude in the groundstate $\ket{\gs}$
for a pair of fermions to be created at top and bottom, connected by a monopole string:
$$ \bra{\gs} f_\text{bottom}  m_2 m_4 ... m_N  f_\text{top}^\dagger\ket{\gs}
= \bra{\gs}   f_\text{top} f_\text{top}^\dagger b_1 b_3 ... b_{N/2} \ket{\gs}
= v^n \bra{\gs}  f_\text{top} f_\text{top}^\dagger  \ket{\gs}
\neq 0 $$
$v^n \sim e^{-L}$ decays exponentially in the thickness of the slab,
but this implies a finite tunneling amplitude.
(Here $n(N) \equiv {N-2 \over 2 } $.)

Since all fermions are charged either electrically or magnetically
(it is ambiguous which should be interpreted as the electron and which as the magnetic monopole),
the fermion number anomaly also implies
a discrete gauge anomaly.  That is,
rotating the phase of every fermion by $\pi$
is part of the $\gU(1)$ gauge group (though not only the electric group in any one duality frame).
This is similar to
Goldstone's understanding of the Witten anomaly
\cite{Goldstone:1984} (as cited in \cite{Elitzur:1984kr, deAlwis:1985uj, Klinkhamer:1990eb}).

Putting two copies of the system together removes the anomaly.
From the point of view above, it is because the monopole strings will reconnect
so that they only attach fermions at the same surface.
A similar mechanism of reconnection was described in \cite{McGreevy:2011if}.

\section{Consequences for all-fermion toric code}
\label{sec:all-fermion-TC}

So far we've discussed
bosonic SPTs in $D=4+1$ with no symmetry, and have briefly mentioned bosonic SPTs in $D=3+1$ with time-reversal ($\CT$) symmetry.
In both cases, there is a symmetry-preserving termination which is a gauge theory where all the matter is fermionic.
There are many illuminating connections between these two problems. To understand them, we must now discuss the $D=3+1$ $\CT$ invariant SPT \cite{PhysRevX.3.011016, PhysRevB.87.174412, PhysRevB.87.235122, Burnell:2013bka, Metlitski:2013uqa} in more detail.

Briefly, the bulk $3+1$ dimensional state is a quantum phase of bosons protected by time reversal symmetry. The bulk theory has a surface termination consisting of $2+1$ dimensional $\IZ_2$ gauge theory in which the charge, the vortex, and charge-vortex composite are all fermions.  As in the case of all-fermion electrodynamics, the statistics of the charge-vortex composite actually follows from those of the charge and the vortex provided there are no gauge-invariant fermions in the spectrum. What does time reversal have to do with such a $2+1$ dimensional state? Naively, the answer is not much: all topological data, e.g., fusion rules, quantum dimensions, braiding phases, etc. are real numbers, so time reversal invariance doesn't seem to provide a constraint on the topological data.

However, there is one piece of topological data which is sensitive to $\CT$ and that is the chiral central charge, $c_-$. Furthermore, in a microscopic bosonic model, the value of $c_-$ is constrained by the topological data.  If we have anyon types labeled by $a$ with quantum dimensions $d_a$ and topological spins $s_a$, then the chiral central charge is determined, $\text{mod}\,8$, by
\cite{Rehren:1989bi, Frohlich:1990ww, K0602}
\be \label{eq:ccformula}
\frac{\sum_a d_a^2 e^{2\pi i s_a}}{\sqrt{\sum_a d_a^2}} = e^{2\pi i c_-/8}.
\ee
In a model of abelian anyons, all $d_a=1$ and the total quantum dimension, $\CD =\sqrt{\sum_a d_a^2}$, is simply the square root of the number of anyon types (including the identity).  The fact that the central charge is only determine $\text{mod}~8$ is not an accident \cite{K0602}. The $\mathsf{E}_8$ state of bosons has no anyonic excitations but has chiral central charge $c_-=8$, hence we may add layers of the $\mathsf{E}_8$ to any anyon model without changing the anyon content but shifting the chiral central charge by $8$.

For the familiar $\IZ_2$ gauge theory in which charges and vortices are bosons, we have $a\in\{1,e,m,em\}$, $d_a = 1$, $s_1=s_e=s_m = 0$, and $s_{em} = 1/2$. Hence \eqref{eq:ccformula} gives 
\be
e^{2\pi i c_-/8} = \frac{3+(-1)}{2} = 1
\ee
hence $c_- = 0 \,\text{mod}\, 8$. In other words, the minimal $\IZ_2$ gauge theory has no chiral edge states. However, if we consider the all-fermion gauge theory, then we find
\beq
e^{2\pi i c_-/8} = \frac{1 + 3(-1)}{2} = -1
\eeq
hence $c_- = 4 \,\text{mod}\, 8$. Thus the all-fermion gauge theory must have chiral edge states and hence must indeed break $\CT$. The reason why this state can be realized
in a $\CT$-invariant manner
at the surface of a $\CT$-invariant $3+1$ bulk state is that in this case it is impossible to create an edge for the gauge theory at which the chiral edge states can be exposed.

Now we turn to connections between the system just discussed
and the all-fermion electrodynamics in $D=3+1$.
First, suppose all-fermion electrodynamics did have a time reversal symmetric bosonic regulator.  Then so does the all-fermion toric code.
The argument is as follows.
Condense pairs of charges in 3+1d (thereby higgsing the gauge group to $\IZ_2$),
and place the system on $\IR^2 \times S^1$, where the radius of the $S^1$ is $L$.
The $\IZ_2$ topological order implies
that
states with different $\IZ_2$ flux through the circle
are split only by an amount of order $E_\text{flux} \sim  e^{ - L | \log t|/ \xi }$
where $t$ is a hopping amplitude for $\IZ_2$ charged quasiparticles,
and $\xi$ is the bulk correlation length.
The regime of interest has $ L \gg \xi$
(so that
our field theory analysis is valid) and
$E_\text{flux} \gg m_e, m_m $,
where $m_e$ and $m_m$ are the rest energies of
the electric and magnetic quasiparticle excitations.
%
The result is then the all-fermionic toric code with, by assumption, a time-reversal symmetric bosonic regulator.
Assuming that no such regulator exists for the all-fermion toric code, no such regulator can exist for all-fermion electrodynamics.
(And as \cite{WangPotterSenthil} point out,
the case with time reversal symmetry is actually the crucial case, in the sense that
the SPTness of the state persists even upon breaking time reversal.)

Second, all-fermion electrodynamics does have a time reversal symmetric {\it fermionic} regulator.
Indeed, it is equivalent to BBF electrodynamics by
binding the neutral fermion to the electron.
(In this case there are particles of both statistics in each charge sector;
for purposes of discussion, we label a model by the statistics of the lightest particle in each sector.)
Again condense charges and compactify on a circle.
This produces a time reversal symmetric fermionic regulator for the all-fermion toric code.  And again, we can convert FFF toric code to BBF toric code in the process.

It is instructive to ask what happens to the chiral central charge formula \eqref{eq:ccformula}. The answer is that the formula only applies when the regulator is bosonic. This is crucial because the $\text{mod}~8$ property of the formula relied on the $\mathsf{E}_8$ phase being the simplest phase with chiral edge states and no anyonic excitations. Once we add microscopic fermions, there are simpler chiral states. The simplest is the $p+ip$ state of fermions with $c_- = 1/2$.  Hence while the minimal chiral central charge, $c_- = 4$, of the all-fermion gauge theory could not be cancelled with only bosonic short-range entangled states (which can only shift $c_-$ by $8$), the minimal central charge of the all-fermion gauge theory can be cancelled by fermionic short-range entangled states (which can shift $c_-$ by $1/2$).

In both cases adding microscopic fermions saves everything, in the sense that
all spectra of excitations are adiabatically connected.

\vfill\eject

{\bf Fermion number anomaly.}
Since the structure of our coupled-layer construction
is so similar to that of
the $D=3+1$ beyond-cohomology boson SPT
in \cite{PhysRevB.87.235122},
the same logic applies to that model
(removing daggers where necessary since
charges are binary).
That is, 
in a slab geometry, 
a state with a fermion on the top surface
can tunnel to a state with a fermion on the bottom surface, 
because 
the quasiparticle sectors are related 
by bosonic operators (some of which are condensed):
$$  f_\text{top} b_1 b_3 ... b_{N/2} = m_2 m_4 ... m_N f_\text{bottom}.$$
We therefore expect that this bosonic state
can transport fermion number between edges.

In this case, the bulk state is protected by time-reversal invariance.
Breaking time reversal only at the surface
produces a state which is still not edgeable. 
We give two examples of time-reversal broken surface states momentarily.
It will help to see the connection between the fermion number anomaly
and the preservation of $\CT$ to ask: 
What happens to the edge if we adiabatically continue the bulk 
through a $\CT$-breaking path to a product state? 
It is not necessary to have a surface phase transition:
Without $\CT$, one way to deform the bulk (on a torus, say) to a product state is to 
open up an array of gapped trivial surfaces (possible because $\CT$ is broken) and then expand the intervening vacuum regions to consume the system, following \cite{2014arXiv1407.8203S}. 
On a system with boundary, this can be done everywhere except at topologically ordered boundaries which are independently stable. 
On a slab of finite but large thickness, therefore,
in the absence of $\CT$, one can disconnect the top from the bottom by cutting open a middle (trivial, gapped) surface, hence ending the fermion tunneling without destroying the surface topological order.

A model with the same spectrum of quasiparticles and braiding statistics
{\it can} be realized intrinsically in $D=2+1$.
For example, it can be obtained from the Kitaev honeycomb model with $ \nu  = 8 $
(see Table 2 of \cite{K0602}).
That model does not preserve time reversal symmetry:
the violations of time-reversal symmetry occur
at boundaries, where there is a chiral edge spectrum (with $c_L - c_R = 4$).
The model at the surface of the boson SPT
cannot be put on a space with boundary
(since the boundary of a boundary is empty)
and is time-reversal invariant.
The price for this extra symmetry is
that the fermion number is not conserved!

To connect the various phenomena, it is useful to explicitly realize various $\CT$ broken surface states starting from the $\CT$ invariant all-fermion surface toric code.  The basic observation follows from the previous paragraph: given $\IZ_2$ charged fermionic matter we may shift the vortex from bosonic to fermionic and vice versa by adding $\nu = \pm 8$ copies of a $p+ip$ state for the charged fermions. Normally in 2+1 dimensions the time-reversal point has an absolute chiral central charge $c_-=0$ and a bosonic vortex.  We can obtain a fermionic vortex and $c_-=4$ by adding $\nu=8$ copies of charged fermions in $p+ip$ states. However, on the surface of the $\CT$ invariant bosonic SPT, there is a shift in the spectrum so that the $\CT$ invariant point has a fermionic vortex.  Then we may construct a pair of $\CT$ broken surface states which are still topologically ordered by adding $\nu = \pm 8$ copies charged fermions in $p+ip$ states.  The system now explicitly breaks time reversal and has a bosonic vortex.

Given a bosonic vortex, we may condense the vortex to destroy the surface topological order.  At a domain wall between the two distinct ways to break $\CT$ to obtain a bosonic vortex we have $\nu = 16$ Majorana edge modes before condensing the vortex. After condensing the vortex we obtain the edge of $\mathsf{E}_8$ state of bosons \cite{Kitaev-unpublished}.  Thus we obtain the same edge physics as the $\mathsf{E}_8$ BF theory discussed in \cite{PhysRevX.3.011016}. This analysis provides another route to connect the layer construction to a topological bulk theory via the non-trivial surface, in this case in $3+1$ dimensions. When the surface preserves $\CT$ we may interpret the bulk $FF$ term in $3+1$ dimensions as providing a $\CT$ invariant regulator for the surface all-fermion toric code.

Again the presence of neutral bulk fermions renders everything trivial.
In the presence of microscopic neutral fermions, the bosonic SPT 
can be deformed into 16 copies of the free fermion topological superconductor,
and this in turn is equivalent to nothing
\cite{Kitaev-2013, 2014arXiv1406.3032M}.
So adding fermions explicitly makes the bulk trivial (in addition to the edge).
This picture nicely complements the edge analysis above where we argued 
that adding fermions effectively changes the minimal chiral central charge one can have without topological order (from $c_- = 8$ to $c_- = 1/2$).

{\bf Reality of this phenomenon.}
We have to ask:
Are there real physical systems made just of bosons, with a gap,
which can transport fermion number?
The $D=3+1$ boson SPT protected by time-reversal should do so.
This makes it even more interesting to try to realize this state in the world.

Finally, we note the following consequence of our claim, 
given that elementary gauge-neutral fermions have not been observed in nature\footnote{Here we mean
`neutral under gauge groups which are unbroken at low energies'; absent 
discrete gauge symmetries, a right-handed neutrino 
would falsify this claim.}.
Were we to discover a fermionic magnetic monopole in our world, it 
would imply either\footnote{We must note some uncertainty involving the role of gravity.}: 
\begin{enumerate}
\item There are microscopic, gauge-neutral fermions.
The opposite is conjectured to be true in {\it e.g.}~Ref.~\cite{wen04}.

%
%
%
\item We live on the boundary of some higher dimensional space. 
Boundary theories of 4+1D SPT phases have been suggested in attempts to understand the matter content 
(and flavor structure) of the standard model
\cite{Kaplan:2009yg, Kaplan:2011vz, Wen:2013ppa, 2014arXiv1402.4151Y}.
\end{enumerate}

\vfill\eject

\vskip.2in
{\bf Acknowledgements.}
We thank C.~Chamon, D.~Das, B.~Dolan, J.~Kaplan, and T.~Senthil for useful discussions and comments. This work was supported in part by
funds provided by the U.S. Department of Energy
(D.O.E.) under cooperative research agreement DE-FG0205ER41360,
in part by the Alfred P. Sloan Foundation.
BGS and JAM acknowledge the hospitality of the Perimeter Institute for Theoretical Physics during
the workshop ``Low Energy Challenges for High Energy Physicists".
Research at Perimeter Institute is supported by the Government of Canada through Industry Canada and by the Province of Ontario through the Ministry of Economic Development \& Innovation.

\appendix

\section{Lattice bosons for duality-symmetric surface QED}

{\bf This is a model of bosons.}
The two-form gauge theory studied in this paper is a model of bosons.
Low-energy evidence for this statement
is the fact that we did not have to choose a spin structure to put it on an arbitrary 4-manifold.
This is in contradistinction to $\gU(1)_{k=1}$ CS theory in $D=2+1$.
We note in passing that on a manifold that admits spinors,
the intersection form is even $(\CI(v,v) \in 2\IZ)$ \cite{Donaldsonbook}.
(This means that to describe an effective field theory
for a {\it fermionic} SPT state, we should consider the level $ k \in \IZ/2$.)

High-energy (\ie~condensed-matter) evidence
for the claim that this is a model of bosons
is the following
conjecture for a lattice model of bosons which produces this EFT.
The Hilbert space is as follows
and is similar to lattice boson constructions of electrodynamics
in other dimensions \cite{Wegner1971, SM0204, LW0510, 2013PhRvB..87d5107V}.

$\bullet$ Put rotors $e^{ \ii b_p}$ on the {\it plaquettes} $p$ of a 4d spatial lattice.
(Actually, the model is defined for any 4d simplicial complex.
Translation invariance will not play a significant role.)
These act as
$$ e^{ \ii b_p} \ket{ n_p} = \ket{n_p+1} $$
on states with definite excitation number $n_p$;
we will interpret $n_p$ as a number of (oriented) `sheets' covering the plaquette.

$\bullet$ Put charge-$k$ bosons $\Phi_\ell = \Phi^\dagger_{-\ell}$ on the {\it links} $\ell$.
These satisfy $ [\Phi_\ell, \Phi_\ell^\dagger] = 1$.
We will say that $\Phi_\ell^\dagger$ creates a string segment,
and $\Phi^\dagger_\ell \Phi_\ell$ is the number of (oriented) strings covering the link.

The Hamiltonian is
\bea \HH =&
-
\underbrace{
\sum_{\text{links}, \ell \in \Delta_1}
( \sum_{p\in s(\ell)}  n_p -  \redt{k}\Phi_\ell^\dagger \Phi_\ell )^2
}
_{\tiny \HH_1,\begin{matrix}
\color{darkblue}\text{gauss law. happy when sheets close,}\cr \text{or end on strings}\end{matrix}}
-& \underbrace{ \sum_{\text{volumes, } v \in\Delta_3}
\prod_{p \in \partial v} e^{ \ii b_p}  + h.c.
}_{ \HH_3 \sim B^2, \color{darkblue}\text{ makes sheets hop.}}
\cr
-&
\nonumber
\underbrace{ \Gamma \sum_{p \in \Delta_2} n_p^2 }_{ \HH_2\text{$\sim E^2$.  \color{darkblue}discourages sheets.} }
-&\underbrace{ t \sum_{p \in \Delta_2}
e^{ \ii \redt{k} b_p}
 \prod_{\ell \in \partial p}   \Phi_\ell^\dagger
 + h.c.
}_{ \HH_\text{strings},~ \color{darkblue}{\text{\tiny hopping term for matter strings}}}
+ V\(|\Phi|^2\)
\eea

The subscripts indicate the dimension of the simplices
to which the terms are associated.
When $\Gamma = 0, V=0$, these terms all commute.
The groundstate for $t>0$ is
described by a soup of oriented closed 2d sheets, groups of $k$ can end on strings.



Now take $V\(|\Phi|^2\) = \( |\Phi|^2 - v^2\) $.
This causes to condense $\Phi_\ell = v e^{\ii \varphi_\ell} $,
which leads to a 2-form higgs mechanism:
$$ \HH_\text{strings}  = - \sum_p t v^4 \cos \( k b_p - \sum_{\ell \in \partial p} \varphi_\ell \) $$
On the low-energy manifold of this Hamiltonian, we have
$$ \( e^{ \ii b_p }\)^k = \Ione , ~~ \ket{n_p} \simeq \ket{n_p +k} .$$
This leaves behind $k$ species of (unoriented) sheets.

The groundstates are then described by equal-superposition sheet soup.
If the intersection form on the spatial 4-manifold
which is triangulated by the simplicial complex has $\CI = \Ione$,
there are $ k^{b_2}$ groundstate  sectors.
These groundstates represent the algebra of `tube operators': for any closed union of 2-simplices $\omega$
$$
\CF_\omega \equiv \prod_{p \in \omega} e^{ \ii b_p }
~~~~~~~~~~~~~
\CT_\omega \equiv \prod_{V \in \Delta_3}
\prod_{p \in \partial V \cap \omega}
n_p $$
$$ \CF_\omega \CT_{\omega'} = e^{ 2\pi \ii  \CI_{\omega \omega'}}  \CT_\omega' \CF_\omega
$$

{\bf Continuum limit.}
The higgs mechanism described above leads to
$U(1) \buildrel{\text{higgs}}\over{\to} \IZ_k$ 2-form gauge theory:
$$ L = {tv^4 \over 2} \( \dd \varphi_1 + k B_2 \) \wedge \star \( \dd \varphi_1 + k B_2 \)
+ {1\over g^2}  \dd B_2 \wedge \star \dd B_2 $$
$$ \simeq
{ k \over 2 \pi }B \wedge \dd C
+ { 1 \over 8 \pi tv^4}
 d C \wedge \star \dd C
 + {1\over g^2}  \dd B \wedge \star \dd B $$
with $ \dd C \simeq 2 \pi t \star (\dd \varphi + k B) $.
This equivalence is described in
\cite{Maldacena:2001ss, 2004AnPhy.313..497H}.


\section{More details on monopole strings and vortex sheets in 5d abelian gauge theory}
\label{app:vortex-sheet}

Consider a 5d U(1) 1-form gauge field $a$, with field strength
$f= da$.
A magnetic excitation with respect to this gauge field has $ \oint_{\Sigma_2} f = 2 \pi g$,
where $\Sigma_2$ is a closed 2-surface {\it surrounding} the object.
Such an object is therefore codimension three, and is a string in 4+1 dimensions.
The quantity
which is localized on the monopole strings is therefore a three-form:
$$\star j_m = \delta^3(\text{monopole strings}) =  \star df \equiv d C$$
where $C$ is a two-form.

Suppose we higgs the $U(1)$ gauge field by condensing a charged
order parameter field $ b \sim  v e^{ i \phi}$.
This adds
$$ \delta H = m^2 (a + d \phi)^2 , $$
so that $a$ eats the phase $\phi$, and $ m \sim t v $.
Topological defects in $\phi$, \ie~zeros of $b$ around which $\phi$ winds by $2 \pi$,
occur at codimension two (since $b$ is a complex function) and in 5d are therefore 2+1-dimensional vortex {\it sheets}.

These vortex sheets can end on the monopole strings.
This is the same fact as the fact that
vortex strings can end on magnetic monopoles in 3+1 dimensions.
In the higgs phase of a 3+1 dimensional abelian gauge theory,
the vortex string provides a means to collimate the magnetic flux coming out of the monopole.
The result is the confinement of the magnetic charges; this is a manifestation of the Meissner effect.
It is the same in $D=4+1$, except now it is magnetically charged {\it strings} which
are connected by vortex {\it sheets}.
In the higgs phase, it is energetically favorable for the monopole strings
to be connected by such vortex sheets.

The final ingredient in the coupled-layer construction
is the fact that the condensate is
not purely electric with respect to any individual layer.


\bibliographystyle{spt}

\bibliography{SPT}

\end{document}